\documentclass{cernyrep}
\usepackage[utf8]{inputenc}
\usepackage[T1]{fontenc}
\usepackage{soul}
\usepackage{xcolor}
\usepackage{hyperref}
\usepackage{mmacells}
\usepackage{bbold}
\usepackage{slashed}
\definecolor{maroon}{cmyk}{0, 0.87, 0.68, 0.32}
\definecolor{halfgray}{gray}{0.55}
\definecolor{ipython_frame}{RGB}{207, 207, 207}
\definecolor{ipython_bg}{RGB}{247, 247, 247}
\definecolor{ipython_red}{RGB}{186, 33, 33}
\definecolor{ipython_green}{RGB}{0, 128, 0}
\definecolor{ipython_cyan}{RGB}{64, 128, 128}
\definecolor{ipython_purple}{RGB}{170, 34, 255}

\usepackage{listings}
\lstdefinelanguage{iPython}{
    morekeywords={access,and,break,class,continue,def,del,elif,else,except,exec,finally,for,from,global,if,import,in,is,lambda,not,or,pass,print,raise,return,try,while},%
    morekeywords=[2]{abs,all,any,basestring,bin,bool,bytearray,callable,chr,classmethod,cmp,compile,complex,delattr,dict,dir,divmod,enumerate,eval,execfile,file,filter,float,format,frozenset,getattr,globals,hasattr,hash,help,hex,id,input,int,isinstance,issubclass,iter,len,list,locals,long,map,max,memoryview,min,next,object,oct,open,ord,pow,property,range,raw_input,reduce,reload,repr,reversed,round,set,setattr,slice,sorted,staticmethod,str,sum,super,tuple,type,unichr,unicode,vars,xrange,zip,apply,buffer,coerce,intern},%
    sensitive=true,%
    morecomment=[l]\#,%
    morestring=[b]',%
    morestring=[b]",%
    morestring=[s]{'''}{'''},
    morestring=[s]{"""}{"""},
    morestring=[s]{r'}{'},
    morestring=[s]{r"}{"},%
    morestring=[s]{r'''}{'''},%
    morestring=[s]{r"""}{"""},%
    morestring=[s]{u'}{'},
    morestring=[s]{u"}{"},%
    morestring=[s]{u'''}{'''},%
    morestring=[s]{u"""}{"""},%
    literate=
    {á}{{\'a}}1 {é}{{\'e}}1 {í}{{\'i}}1 {ó}{{\'o}}1 {ú}{{\'u}}1
    {Á}{{\'A}}1 {É}{{\'E}}1 {Í}{{\'I}}1 {Ó}{{\'O}}1 {Ú}{{\'U}}1
    {à}{{\`a}}1 {è}{{\`e}}1 {ì}{{\`i}}1 {ò}{{\`o}}1 {ù}{{\`u}}1
    {À}{{\`A}}1 {È}{{\'E}}1 {Ì}{{\`I}}1 {Ò}{{\`O}}1 {Ù}{{\`U}}1
    {ä}{{\"a}}1 {ë}{{\"e}}1 {ï}{{\"i}}1 {ö}{{\"o}}1 {ü}{{\"u}}1
    {Ä}{{\"A}}1 {Ë}{{\"E}}1 {Ï}{{\"I}}1 {Ö}{{\"O}}1 {Ü}{{\"U}}1
    {â}{{\^a}}1 {ê}{{\^e}}1 {î}{{\^i}}1 {ô}{{\^o}}1 {û}{{\^u}}1
    {Â}{{\^A}}1 {Ê}{{\^E}}1 {Î}{{\^I}}1 {Ô}{{\^O}}1 {Û}{{\^U}}1
    {œ}{{\oe}}1 {Œ}{{\OE}}1 {æ}{{\ae}}1 {Æ}{{\AE}}1 {ß}{{\ss}}1
    {ç}{{\c c}}1 {Ç}{{\c C}}1 {ø}{{\o}}1 {å}{{\r a}}1 {Å}{{\r A}}1
    {€}{{\EUR}}1 {£}{{\pounds}}1
    {^}{{{\color{ipython_purple}\^{}}}}1
    {=}{{{\color{ipython_purple}=}}}1
    {+}{{{\color{ipython_purple}+}}}1
    {*}{{{\color{ipython_purple}$^\ast$}}}1
    {/}{{{\color{ipython_purple}/}}}1
    {+=}{{{+=}}}1
    {-=}{{{-=}}}1
    {*=}{{{$^\ast$=}}}1
    {/=}{{{/=}}}1,
    literate=
    *{-}{{{\color{ipython_purple}-}}}1
     {?}{{{\color{ipython_purple}?}}}1,
    identifierstyle=\color{black}\ttfamily,
    commentstyle=\color{ipython_cyan}\ttfamily,
    stringstyle=\color{ipython_red}\ttfamily,
    keepspaces=true,
    showspaces=false,
    showstringspaces=false,
    rulecolor=\color{ipython_frame},
    frame=single,
    frameround={t}{t}{t}{t},
    framexleftmargin=6mm,
    numbers=left,
    numberstyle=\tiny\color{halfgray},
    backgroundcolor=\color{ipython_bg},
    basicstyle=\footnotesize\ttfamily,
    keywordstyle=\color{ipython_green}\ttfamily,
    aboveskip=1.2em,
    belowskip=1.2em,
}

\usepackage{listings}
\usepackage{fontawesome}
\usepackage[most]{tcolorbox}
\definecolor{light-gray}{gray}{0.9}

\hypersetup{
  linkcolor  = black, 
  citecolor  = blue,
  urlcolor   = blue,
  colorlinks = true,
  breaklinks=true
}

\graphicspath{{./Figures/}{./}}

\newcommand{\sscript}[1]{{\scriptscriptstyle \mathrm{#1}}}
\newcommand{\rep}[1]{\mathbf{#1}}

\newcommand{\LL}{\mathrm{L}}

\newcommand{\SU}{\mathrm{SU}}

\newcommand{\Psl}{\slashed{P}}
\newcommand{\STr}{\text{STr}}
\newcommand{\STrEAM}{\texttt{STrEAM}}

\definecolor{codegris}{rgb}{0.92,0.92,0.92}

\begin{document}

\preprint{CERN-LHCEFTWG-2022-002\\CERN-LPCC-2022-07}
\date{November, 2020}

\title{LHC EFT WG Note:\\ Precision matching of microscopic physics to the Standard Model Effective Field Theory (SMEFT)}

\author{
Sally Dawson\twoaff{1}{a}, 
Admir Greljo\threeaff{2}{15}{a},
Kristin Lohwasser\twoaff{3}{a},
\\
Jason Aebischer\aff{4},
Supratim Das Bakshi\aff{5},
Adri\'an Carmona\aff{5},
Joydeep Chakrabortty\aff{6},
Timothy Cohen\aff{7},
Juan Carlos Criado\aff{8},
Javier Fuentes-Mart\'in\aff{5},
Achilleas Lazopoulos\aff{9},
Xiaochuan Lu\aff{7},
Stefano Di Noi\twoaff{12}{13},
Pablo Olgoso\aff{5}, 
Sunando Kumar Patra\aff{10},
Jos\'e Santiago\aff{5},
Luca Silvestrini\aff{14},
Anders Eller Thomsen\aff{2},
Zhengkang Zhang\aff{11}
}

\institute{
\naff{1}{Department of Physics, Brookhaven National Laboratory, Upton, NY, United States}
\naff{2}{Albert Einstein Center for Fundamental Physics, Institute for Theoretical Physics, University of Bern, Bern, Switzerland}
\naff{3}{Department of Physics and Astronomy, University of Sheffield, Sheffield, United Kingdom}
\naff{4}{Physik-Institut, Universität Zürich, Zürich, Switzerland}
\naff{5}{Departmento Fisica Teorica y del Cosmos, Universidad de Granada, Granada, Spain}
\naff{6}{Department of Physics, Indian Institute of Technology, Kanpur, 208016, India}
\naff{7}{Institute for Fundamental Science, University of Oregon, Eugene, OR, United States}
\naff{8}{Department of Physics, Durham University, Durham, United Kingdom}
\naff{9}{Institute for Theoretical Physics, ETH Zürich, Zürich, Switzerland}
\naff{10}{Bangabasi Evening College, Kolkata, India}
\naff{11}{Department of Physics, University of California, Santa Barbara, CA, United States}
\naff{12}{Dipartimento di Fisica e Astronomia “G. Galilei”, Università degli Studi di Padova, Padua, Italy}
\naff{13}{Istituto Nazionale di Fisica Nucleare, Sezione di Padova, Padua, Italy}
\naff{14}{Istituto Nazionale di Fisica Nucleare, Sezione di Roma, Rome,
Italy}
\naff{15}{Department of Physics, University of Basel, Basel, Switzerland}
\naff{a}{Convenors of the LHC EFT working group area 5}
}

\begin{abstract}
This note gives an overview of the tools for the precision matching of ultraviolet theories to the Standard Model effective field theory (SMEFT) at the tree level and one loop. Several semi- and fully automated codes are presented, as well as some supplementary codes for the basis conversion and the subsequent running and matching at low energies. A suggestion to collect information for cross-validations of current and future codes is made.
\end{abstract}

\keywords{EFT, One-Loop Matching, Running, UV models}

\maketitle

\tableofcontents

\section{Introduction and Motivation}

The Standard Model effective field theory (SMEFT) describes physics at energies below the new mass scale which is assumed to be above the electroweak scale. The imprints of ultraviolet (UV) physics are encoded in the Wilson coefficients (WC) of the SMEFT. Measuring these coefficients and their correlations allows for discriminating between different UV models. The important technical step in this procedure is the \textit{matching}, where the heavy degrees of freedom are integrated out and their effects are represented by local operators. The resulting WC are expressed in terms of the parameters of the UV theory such as couplings and masses. This facilitates the interpretation of the SMEFT analyses in explicit UV models. 

Matching beyond the tree level is important since many interesting observables are generated only at the one-loop level. 
However, this task is not only technically challenging but given the number of possible UV models, repetitive and time-consuming. To address the issue, several dedicated tools have been developed recently. For example, the {\tt SuperTracer}~\cite{Fuentes-Martin:2020udw}, {\tt Matchete} (to be released), {\tt STrEAM}~\cite{Cohen:2020qvb} and \texttt{CoDEx} \cite{Bakshi:2018ics} packages aim at facilitating the one-loop EFT matching of generic UV models using path-integral methods. {\tt Matchmakereft}~\cite{Carmona:2021xtq}, instead, automates the diagrammatic EFT matching of generic UV models. These tools are introduced in Sec.~\ref{sec:Codes} where also possible avenues for code validation and benchmarking are described. 

Furthermore, there are several codes on the market that deal with Renormalization Group Evolution (RGE) and the treatment of numerical Wilson coefficient values. Such tools are especially important in phenomenological analyses but also when comparing analytic matching results obtained from the above-mentioned matching codes. In Sec.~\ref{sec:NumCodes} some of these numerical tools are discussed. Namely, the (match)runner codes \texttt{DsixTools}~\cite{Celis:2017hod,Fuentes-Martin:2020zaz}, \texttt{RGESolver}~\cite{DiNoi:2022ejg} and \texttt{wilson}~\cite{Aebischer:2018bkb}, and the Wilson coefficient exchange format (\texttt{WCxf})~\cite{Aebischer:2017ugx}.

\section{Matching Codes}
\label{sec:Codes}

Codes to (semi-)automatically match a concrete UV model to the SMEFT are important tools for constraining beyond the Standard Model (BSM) theories by the global SMEFT fits. An overview of the different codes and their primary functions is given below. The codes are introduced according to strict alphabetical order. Users should therefore study the full list before deciding which code is best suited for their needs.

\subsection{{\tt CoDEx}}

\texttt{CoDEx} \cite{Bakshi:2018ics} is a Mathematica \cite{Mathematica} package that integrates out heavy fields of spin-$0,1/2,1$ and computes  the  effective operators up to mass dimension-6 and associated Wilson coefficients (WCs) in terms of the model parameters. Relying on the functional method, it can perform the integration out at both tree- and 1-loop-levels. It offers the effective operators in both \texttt{SILH} \cite{Giudice:2007fh,Elias-Miro:2013mua} and \texttt{Warsaw} \cite{Buchmuller:1985jz,Grzadkowski:2010es} bases. \texttt{CoDEx} can deal with BSM scenarios containing single or multiple mass-degenerate heavy fields of the same spin. To run the program, it requires very minimal input within a user-friendly format. The user needs to provide only the relevant part of the BSM Lagrangian that involves the heavy field(s) to be integrated out.  \texttt{CoDEx} generates the effective action using functional method \cite{Gaillard:1985uh,Chan1985,Cheyette:1987qz,Bilenky:1993bt,Henning:2014wua,Drozd:2015rsp,Kramer:2019fwz,Angelescu:2020yzf,Fuentes-Martin:2020udw,Cohen:2020qvb,Fuentes-Martin:2016uol}, and an internal program is used to identify the effective operators and accompanying WCs. The operators are computed at the energy scale where the integration out is performed, i.e.,  the mass of the heavy field(s).   \texttt{CoDEx} provides an option to invoke the RGE of the effective operators in Warsaw basis using the anomalous dimension matrices \cite{Jenkins:2013zja,Jenkins:2013wua,Alonso:2013hga} and note down the set of operators that emerge at any other scale. \texttt{CoDEx} with its installation instructions, web documentation, and model examples is available on GitHub. \href{https://effexteam.github.io/CoDEx/}{\color{red}\faGithub}.

 \

\begin{figure}[ht]
		\centering
		\includegraphics[width=16cm,height=10cm]{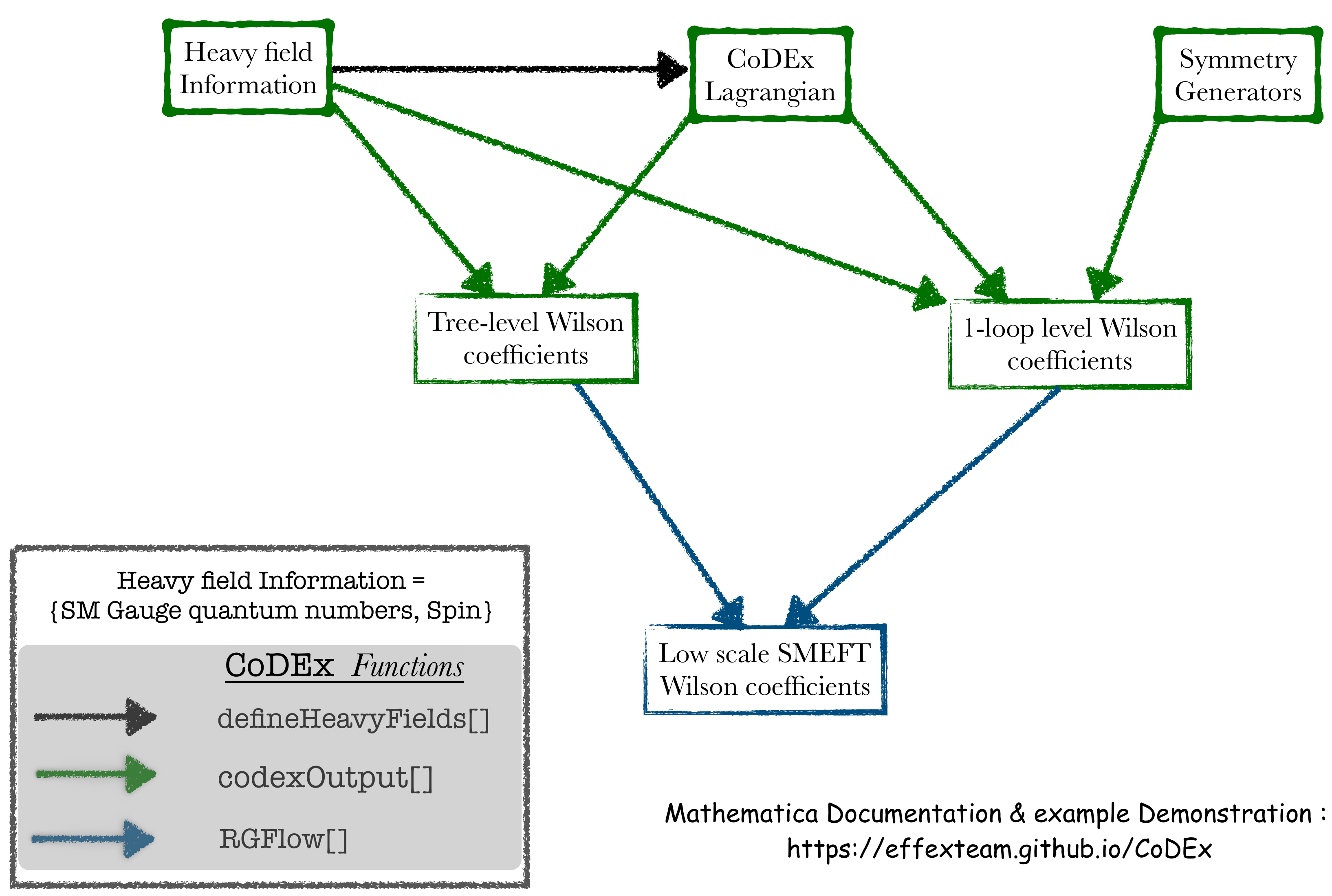}
		\caption{Flow-chart for \texttt{CoDEx}. The inputs and outputs of the \texttt{CoDEx}-functions are represented by coloured lines.}
		\label{fig:flowch}
	\end{figure}
	
	\
	
{\bf \underline{User Inputs \& Possible Outputs}}

	\indent The input information to run \texttt{CoDEx} for any given BSM scenario is minimal. Here, we depict a step-by-step procedure to compute the effective operators and the internal computation that is carried out at each step in \texttt{CoDEx}:
	
	\begin{itemize}
		\item User needs to provide the following information about the heavy field(s): Color, Isospin, Hyper-charge, Mass, and Spin, based on which the representation(s) of the heavy field(s) are evaluated by the package internally. On top of that, the relevant part of the BSM Lagrangian that involves the heavy field(s) must be supplied by the user.
		
		\item Relying on these inputs, the package can integrate out propagators from the tree- and 1-loop-level processes of that BSM theory. The derivative term and the mass term of the heavy field are internally constructed in the package with the help of quantum numbers provided by the user.
		
		\item The \texttt{CoDEx}-function\,-- \mmaInlineCell{Code}{treeOutput} integrates out the heavy tree-level propagators only and generates the tree-level WCs and effective operators. 
		Internally, \texttt{CoDEx} computes the heavy field classical solution by solving the Euler-Lagrange equation upto an order that will contribute to the dimension-6 operators. Then, the solution is substituted back in the BSM Lagrangian to generate the effective Lagrangian. After implementing appropriate mass-dimension cuts and operator identities to match with the given SMEFT basis, the desired output is generated.
		
		\item The \texttt{CoDEx}-function\,-- \mmaInlineCell{Code}{loopOutput} integrates out heavy propagators from loops and generates the 1-loop-level WCs and effective operators. In addition to the inputs required to generate tree-level operators, \mmaInlineCell{Code}{loopOutput} needs the symmetry generators of the gauge groups, under which the heavy field is charged. The symmetry generators of the SM gauge groups for frequently used representation are available in \texttt{CoDEx}, and thus the user does not need to provide that information externally. To compute the 1-loop generated effective operators, \texttt{CoDEx} internally computes the trace of the effective action formulae \cite{Henning:2014wua,Henning:2016lyp,Zhang:2016pja}. The package recognizes the terms quadratic in the heavy field from the BSM Lagrangian. It builds the covariant derivative operator using the quantum numbers of the heavy field.

		\item Following the above steps, the user generates the effective operators at the matching scale. But one can also find the operators at the electroweak or other suitable scale using \texttt{CoDEx}-function:\mmaInlineCell{Code}{RGFlow}.  This function computes RGE for \texttt{Warsaw} basis effective operators using anomalous dimension matrices available in Refs.~\cite{Jenkins:2013wua,Jenkins:2013zja,Alonso:2013hga}.\vspace{2em}
	\end{itemize}
	
	{\bf \underline{Developers' version: yet to be released}}
	\begin{itemize}
		\item {\bf WCxF \cite{Aebischer:2017ugx}:} We have added two \texttt{CoDEx}-functions: \mmaInlineCell{Code}{wcxfOut} and \mmaInlineCell{Code}{wcxfIn} to export and import the WCs in a format compatible with other codes, see Refs.~\cite{Proceedings:2019rnh}. There exist several packages with different EFT utilities (facilitating WC matching, renormalisation group running, and calculating observables). However, not all of them are on the same footing, e.g., they use different EFT bases and operator normalization. It is important to have a data exchange interface among these programs, and WCxF provides that. Thus,  it is desirable for any package to have import \& export functions in this format to interface the program with others.
		
		\item {\bf Heavy-light mixed WCs:} The mixed heavy-light contribution, for scalars only, is included in the	matching result by expanding the UV action around the light field solution obtained using the Euler-Lagrange equation, similar to the pure heavy-loop approach. Using the ‘covariant diagrams’ methodology presented in Ref.~\cite{Zhang:2016pja}, we have calculated the formula for the mixed heavy-light contributions and cross-checked it with that of Ref.~\cite{Ellis:2017jns} (see Tables 1–5 in there). We implement this formula in \texttt{CoDEx} along with the 16 BSMs to generate the mixed heavy-light Wilson coefficients \cite{Bakshi:2020eyg,Anisha:2020ggj,Anisha:2021hgc}.
		
		\item {\bf Identities:} Evaluating the effective action for a UV model may generate gauge-invariant structures which do not directly resemble the desired effective operator basis. Then, we require the implementation of operator identities and equations of motion (EOMs) of light degrees of freedom on the effective Lagrangian to transform the gauge-invariant terms to desired structures. The implementation of these identities depends upon the choice of the effective operator basis. These transformations like Fierz identities, SM fields' equations of motion, and SMEFT dimension-6 operator identities are introduced in the developer version of \texttt{CoDEx}. These transformations are necessary to represent the effective Lagrangian in terms of the SMEFT operators and their WCs.
	\end{itemize}

\subsection{{\tt Matchete} and {\tt SuperTracer}}

\begin{figure}
    \centering
    \includegraphics[width= .9\textwidth]{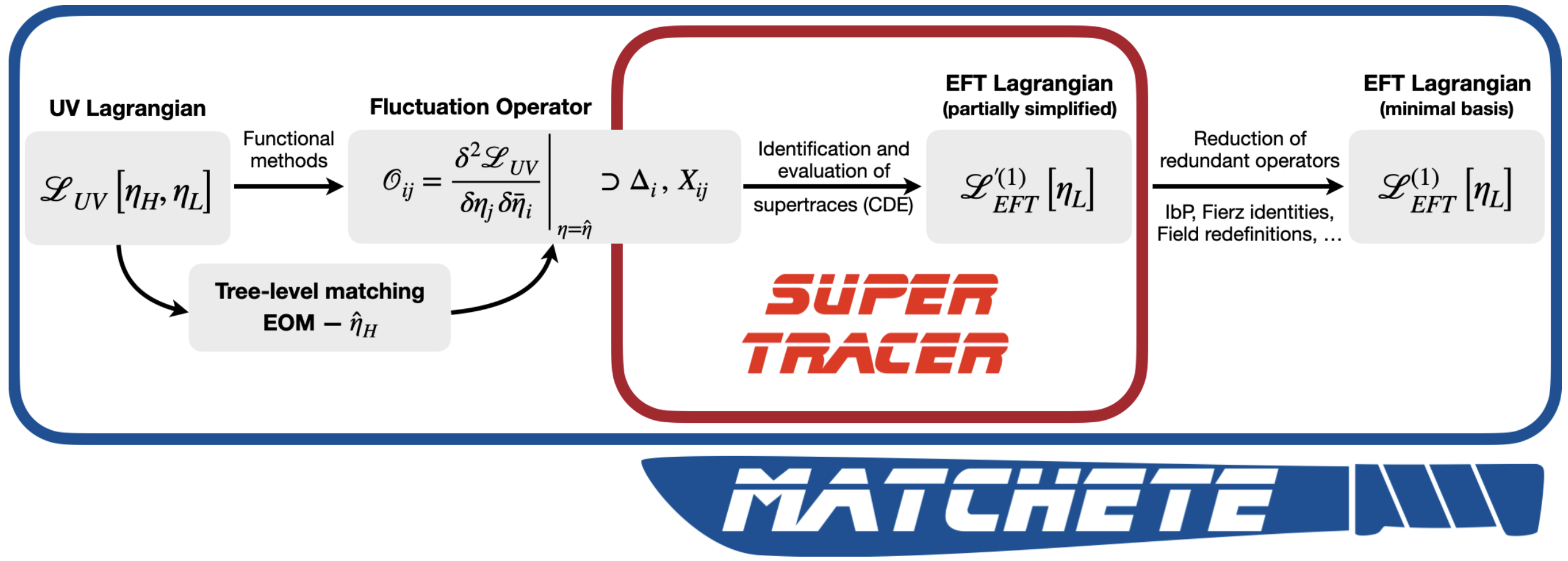}
    \caption{Workflow for functional one-loop matching in the {\tt Matchete} and {\tt SuperTracer} packages.}
    \label{fig:matchete_workflow}
\end{figure}

\texttt{Matchete} and \texttt{SuperTracer} are Mathematica packages aimed at automating the complete one-loop matching of arbitrary UV models into their EFTs, using the functional-matching procedure described in~\cite{Fuentes-Martin:2020udw}. The workflow of these packages is summarized in Fig.~\ref{fig:matchete_workflow}. \texttt{SuperTracer} allows for the evaluation of generic supertraces, one of the most time-consuming and repetitive tasks at the center of functional matching computations. In the future, \texttt{Matchete} is planned to supersede \texttt{SuperTracer} and provide a comprehensive and fully automated matching tool, with a user-friendly interface that will only require the UV Lagrangian as user input.  A proof of concept for \texttt{Matchete} will be made publicly available soon~\cite{Matchete}.

The functional one-loop matching procedure is performed by evaluating the hard region~\cite{Fuentes-Martin:2016uol} of two types of functional supertraces, log-type, and power-type supertraces, corresponding respectively to the first and second term in the following expression:
\begin{equation}\label{eq:STMaster}
    S_\sscript{EFT}^{(1)} = \frac{i}{2} \mathrm{STr} \ln \Delta^{-1} \Big|_{\mathrm{hard}} - \frac{i}{2} \sum_{n=1}^\infty \mathrm{STr} [(\Delta  X)^n ]\Big|_{\mathrm{hard}},
\end{equation}
where $S_\sscript{EFT}^{(1)}$ is the one-loop EFT action, $\Delta $ is the gauge-invariant kinetic operator, and $ X $ the interaction terms. These can be derived directly from the UV Lagrangian by  
    \begin{equation}
    \dfrac{\delta^2 \mathcal{L}_\sscript{UV}}{\delta \eta_j \delta \bar{\eta}_i} = \delta_{ij} \Delta_i^{-1} - X_{ij}.
    \end{equation}
In this formalism, $ \eta_i $ runs over all fields of the theory (counting also conjugate fields). The calculation of the functional supertraces is kept explicitly gauge invariant by doing a Covariant Derivative Expansion (CDE)~\cite{Gaillard:1985uh,Chan:1986jq,Cheyette:1987qz} of both propagators and interaction terms. 

\texttt{Supertracer} can perform the CDE and loop integration of the supertraces for arbitrary interaction terms to get generic expressions for the one-loop EFT. To install the package simply run the following command in a new Mathematica notebook:
\begin{mmaCell}{Input}
Import["https://gitlab.com/supertracer/supertracer/-/raw/master/install.m"]
\end{mmaCell}
This will download and install \texttt{Supertracer} in the Applications folder in the base directory of Mathematica. After the package has been installed, it can be loaded into a Mathematica kernel with the command
\begin{mmaCell}{Input}
<< \mmaDef{SuperTracer\(\,\grave\,\)}
\end{mmaCell}

\texttt{Supertracer} operates with an expansion in light mass dimensions, which provides an expansion parameter for the power-type supertraces and for the CDE. The kinematics of the supertraces change depending on the spin and masses of the propagating fields. Accordingly, \texttt{Supertracer} distinguishes between heavy and light scalars, fermions, vectors, and ghost fields, denoted as \mmaInlineCell[]{Input}{\(\Phi,\phi,\Psi,\psi\),\mmaDef{V},\mmaDef{A},\mmaDef{cV},\mmaDef{cA}}, respectively.
The generic form of the log-type supertrace for heavy fermions up to dimension 6, is then found with the routine 
\begin{mmaCell}{Input}
  \mmaDef{LogTerm}[\(\Psi\),6]
\end{mmaCell}
\begin{mmaCell}{Output}
  -\mmaFrac{1}{6} Log\(\Big[\)\mmaFrac{\mmaSup{\(\overline{\mu}\)}{2}}{\mmaSubSup{M}{H}{2}}\(\Big]\)\mmaSup{G}{\(\mu\nu\)}** \mmaSup{G}{\(\mu\nu\)} + \mmaFrac{1}{15}\mmaFrac{1}{\mmaSubSup{M}{H}{2}}\mmaSub{D}{\(\mu\)}\mmaSup{G}{\(\mu\nu\)}** \mmaSub{D}{\(\rho\)}\mmaSup{G}{\(\nu\rho\)} + \mmaFrac{1}{90}i\mmaFrac{1}{\mmaSubSup{M}{H}{2}}\mmaSup{G}{\(\mu\nu\)}** \mmaSup{G}{\(\mu\rho\)}** \mmaSup{G}{\(\nu\rho\)}
\end{mmaCell} 
Power-type supertraces depend on the $X$ terms that are involved in the trace. To extract the generic form, one has to put in a list of $ X $ terms defining the types of the propagating fields in the loop, and the light dimension of each $X$. Thus, to extract the trace with a heavy fermion and a gauge field with interaction terms of dimension $5/2$ (the smallest light dimension of a heavy fermion field), up to dimension 6, we call
\begin{mmaCell}{Input}
  \mmaDef{STrTerm}[\{\mmaDef{X}[\{\(\Psi\),\mmaDef{A}\}, 5/2],\mmaDef{X}[\{\mmaDef{A},\(\Psi\)\}, 5/2]\}, 6]
\end{mmaCell}
\begin{mmaCell}{Output}
  \mmaFrac{1}{8}i\(\bigg(\)3 + 2 Log\(\Big[\)\mmaFrac{\mmaSup{\(\overline{\mu}\)}{2}}{\mmaSubSup{M}{H}{2}}\(\Big]\)\(\bigg)\)\mmaSub{\(\gamma\)}{\(\mu\)}**\mmaSub{D}{\(\mu\)}\mmaSubSup{X}{\mmaSub{\(\Psi\)}{i}\mmaSub{A}{j}} **\mmaSubSup{X}{\mmaSub{\mmaSub{A}{j}\(\Psi\)}{i}} +\mmaFrac{1}{2}\(\bigg(\)1 + Log\(\Big[\)\mmaFrac{\mmaSup{\(\overline{\mu}\)}{2}}{\mmaSubSup{M}{H}{2}}\(\Big]\)\(\bigg)\)\mmaSub{M}{H}\mmaSubSup{X}{\mmaSub{\(\Psi\)}{i}\mmaSub{A}{j}} **\mmaSubSup{X}{\mmaSub{\mmaSub{A}{j}\(\Psi\)}{i}}
\end{mmaCell}

With the generic formulas for the supertraces, there is limited possibility for simplifications of the expressions, as e.g. contractions in the Dirac algebra and covariant derivatives cannot be fully resolved. Furthermore, in more realistic models, the $ X $ terms, being matrices in field space, are large objects, and expanding the traces by hand is a huge effort.
\texttt{Supertracer} provides the option of directly substituting explicit expressions for the $X$ terms into the traces, which allows for the internal routines to simplify the results as much as possible. This makes it feasible to use \texttt{SuperTracer} for realistic BSM matching computations. 

For proper handling of indices and to include the action of the gauge fields on the various matter fields, the user has to define various objects to properly perform the substitution of the $X$ terms. One has to be rather careful when doing this, and we refer the user to the full manual~\cite{Fuentes-Martin:2020udw}. Here we restrict ourselves to a simple model where, as in the previous supertrace, the interactions are between a heavy fermion and an Abelian gauge field:
\begin{mmaCell}{Input}
  \mmaDef{STrTerm}[\{\mmaDef{X}[\{\(\Psi\),\mmaDef{A}\}, 5/2], \mmaDef{X}[\{\mmaDef{A},\(\Psi\)\}, 5/2]\}, 6,
     \{
      \{\(\Psi\),\mmaDef{A}\}->\{\{-e \(\gamma\)[\mmaUnd{\(\alpha\)}[j]]**\mmaDef{\(\psi\)h[]}\}, \{e \(\gamma\)[\mmaUnd{\(\alpha\)}[j]]**\mmaDef{CConj}[\mmaDef{\(\psi\)h[]]}\}\},
      \{\mmaDef{A},\(\Psi\)\}->\{\{-e \mmaDef{Bar}[\mmaDef{\(\psi\)h}[]]**\(\gamma\)[\mmaUnd{\(\alpha\)}[i]], e \mmaDef{Bar}[\mmaDef{CConj}[\mmaDef{\(\psi\)h[]]}]**\(\gamma\)[\mmaUnd{\(\alpha\)}[i]]\}\},
      \mmaDef{M}[\(\Psi\)]->\{\mmaUnd{Mh}, \mmaUnd{Mh}\},
      \mmaDef{G}[\(\Psi\)]->\{\{e[1]\},\{[e[-1]]\}\},
      \mmaDef{G}[\mmaDef{A}]->\{\{\}\}
     \}
   ]
\end{mmaCell}
\begin{mmaCell}{Output}
  \(\,\)\mmaFrac{1}{2}\textbf{}i\(\,\)\mmaSup{e}{2}\(\bigg(\)1 + 2 Log\(\Big[\)\mmaFrac{\mmaSup{\(\overline{\mu}\)}{2}}{\mmaSup{Mh}{2}}\(\Big]\)\(\bigg)\)\mmaDef{\(\overline{\psi\mathrm{h}}\)}**\mmaSub{\(\gamma\)}{\(\mu\)}**\mmaSub{D}{\(\mu\)}\(\psi\mathrm{h}\) - 2\(\,\)\mmaSup{e}{2}\(\,\)Mh\(\bigg(\)1 + 2 Log\(\Big[\)\mmaFrac{\mmaSup{\(\overline{\mu}\)}{2}}{\mmaSup{Mh}{2}}\(\Big]\)\(\bigg)\)\mmaDef{\(\overline{\psi\mathrm{h}}\)}**\mmaDef{\(\psi\)h}
\end{mmaCell}

The design goal of \texttt{Matchete} is to completely automate the implementation of functional matching. This will include a simple user interface for the UV Lagrangian input, and the implementation of functional derivatives to automate the computation of fluctuation operators and the solution of the EOMs for the heavy fields. For the user, this will result in a tremendous simplification, as the input is directly on the Lagrangian level, rather than having to provide the much more complicated functional objects. The second pillar of \texttt{Matchete} is the integration of robust simplification routines utilizing integration-by-parts (IBP) identities, Fierzing, and field redefinitions to bring the output of the supertraces to an operator basis. Whereas \texttt{SuperTracer} requires the user to be familiar with functional matching, \texttt{Matchete} will be approachable even with rudimentary knowledge.

On a practical level, both \texttt{SuperTracer} and \texttt{Matchete} use dimensional regularization with $\overline{\mathrm{MS}}$ renormalization, with $\gamma_5$ being treated in the naive dimensional regularization (NDR) scheme. Gauge invariance of the resulting EFT is ensured by performing the matching computation in the background field gauge.
Furthermore, the metric signature is chosen to be $ g_{\mu\nu} = (+,\, -,\, -,\,-)$ and the Levi-Civita tensor convention to be $ \varepsilon^{0123} = +1$ .

\subsection{{\tt Matchmakereft}}\label{sec:MMEFT}

\texttt{Matchmakereft} is a Python tool to perform the matching of arbitrary models onto arbitrary effective theories up to one-loop order in an automated way. The matching is performed in a diagrammatic fashion by matching one-light-particle-irreducible (1LPI) off-shell amplitudes functions in the background field method. We currently use dimensional regularization with $\overline{\rm MS}$ renormalization and use an anti-commuting $\gamma^5$ convention, which is enough for most of the relevant calculations in the SMEFT and its simple generalizations.  Further information can be found in the project web page~\url{https://ftae.ugr.es/matchmakereft/} and in the manual~\cite{Carmona:2021xtq}. Its code is publicly available at \url{https://gitlab.com/m4103/matchmaker-eft} and it can be installed via \texttt{PyPI}
\begin{verbatim}
    > python3 -m pip install matchmakereft
\end{verbatim}
or \texttt{Conda}
\begin{verbatim}
    > conda install -c matchmakers matchmakereft
\end{verbatim}
\texttt{Matchmakereft} uses well-tested tools for the different parts of the matching calculation and has therefore some dependencies that have to be met by the user. These include Mathematica, \texttt{FeynRules}~\cite{Alloul:2013bka}, \texttt{FORM}~\cite{Ruijl:2017dtg} and \texttt{QGRAF}~\cite{Nogueira:1991ex} (see the manual for details on how to install them).

Models (both UV and effective) are defined via the Mathematica package \texttt{FeynRules} following the standard rules of that package. One particularity is that every particle has to be defined as heavy or light by using the following rule in their definition
\verb+FullName -> "light"+ for light particles and \verb+FullName -> "heavy"+ for heavy ones. This defines models as \textit{light} models when no heavy particles are present, and \textit{heavy} models, when there are some heavy particles present. EFT models have to be light models whereas UV models can be heavy or light. In the former case, the finite tree-level and one-loop matching is performed, in the latter, the one-loop anomalous dimensions are computed. Depending on the type of model and its properties, further information has to be provided. This includes gauge information, possible symmetry properties of the different couplings, and the explicit reduction from a Green basis to a physical one in the case of the EFT. This latter point is very important. Given that \texttt{Matchmakereft} performs the matching off-shell it is essential that a full Green basis is defined as the EFT. Similarly, the use of the background field method is crucial for the matching to be gauge-independent. Given that the model definition is the only step of the process that is not fully automated and therefore more error-prone, we encourage the reader to consult the manual for all the details.

Once installed, \texttt{Matchmakereft} can be started by typing \verb+matchmakereft+ in the terminal. This loads the CLI that looks like this
\begin{lstlisting}[language=iPython]
Checking for updates.
matchmakereft is up-to-date.

Welcome to matchmakereft v1.0.2
Please refer to arXiv:2112.10787 when using this code. 

matchmakereft> 
\end{lstlisting}
The following commands are currently available:
\begin{itemize}
    \item 
    \begin{lstlisting}[language=iPython]
    matchmakereft> test_installation\end{lstlisting}
    This command tests the installation by running three sample calculations and comparing the results against the known ones in the literature.
    \item \begin{lstlisting}[language=iPython]
    matchmakereft>copy_models modellocation\end{lstlisting}
    This command creates a directory \texttt{MatchMakerEFT} under \texttt{modellocation} with some sample models that can be used as starting points for further model generation. In particular the B-preserving SMEFT model is provided.
    \item \begin{lstlisting}[language=iPython]
    matchmakereft>create_model modefile1.fr ... modefilen.fr\end{lstlisting}
    This command creates a \texttt{Matchmakereft} model under directory \verb+modefilen_MM+. Some extra files could be needed for model creation. Please check the manual for details.
    \item \begin{lstlisting}[language=iPython]
    matchmakereft>match_model_to_eft UVModelName EFTModelName\end{lstlisting}
    This command computes the hard-region contribution to all the required amplitudes to match the UV model under directory \texttt{UVModelName} onto the EFT under directory \texttt{EFTModelName} and compares the two calculations to perform the matching. The corresponding matching is stored in \texttt{UVModelName/MatchingResult.dat} at three different levels of the calculation. First the matching in the Green basis, then the same one after canonical normalization and finally the matching in the physical basis (provided the corresponding reduction was provided during model generation). The matching of the gauge couplings in the background field method is also provided. A significant number of cross checks are performed using the redundancy inherent to the off-shell matching in the background field method. If any inconsistency is found, it is reported with some details stored in the file \texttt{UVModelName/MatchingProblems.dat}.
    \item \begin{lstlisting}[language=iPython]
    matchmakereft>compute_rge_model_to_eft UVModelName EFTModelName\end{lstlisting}
    This command computes the one-loop anomalous dimensions of the Wilson coefficients of the EFT under directory  \verb+EFTModelName+ assuming the (also light) UV model under directory \verb+UVModelName+. The corresponding anomalous dimensions are stored in \verb+UVModelName/RGEResult.dat+.
    \item \begin{lstlisting}[language=iPython]
    matchmakereft>clean_model Model\end{lstlisting}
    This command restarts model \texttt{Model} to repeat the matching calculation from scratch.
    \item \begin{lstlisting}[language=iPython]
    matchmakereft>check_linear_dependence EFTModelName\end{lstlisting}
    This command checks if the operators defined in EFTModelName are (off-shell) linearly independent or not and if they are not, it provides the linear relations among them. 
    \end{itemize}
In order to add flexibility, \texttt{Matchmakereft} adds some commands to split the calculation of the matching in smaller steps (amplitude calculation and Wilson coefficient calculation) and also includes the possibility of performing only tree-level matching. Further details can be found in the manual.

\texttt{Matchmakereft} is in very active development and we encourage the user to always update to the latest version and to check the manual or the web page for updates on its functionality.

\subsection{{\tt MatchingTools}}

\texttt{MatchingTools}~\cite{Criado:2017khh} is a Python package for performing tree-level matching calculations between general EFTs, and for implementing the algebraic manipulations of effective Lagrangians needed to re-write them in terms of a basis of operators. Its code is publicly available at \href{https://github.com/jccriado/matchingtools}{github.com/jccriado/matchingtools}, and it can be installed through:
\begin{verbatim}
  > pip3 install matchingtools
\end{verbatim}
The main focus of \texttt{MatchingTools} is on applications related to the SMEFT, but it works in a more general setting. It can integrate out fields of spin 0, 1/2 or 1 out of the box. Fields of higher spin can also be included by providing the corresponding propagators. No assumptions are made about their transformation properties under internal symmetry groups. Regarding their interactions, the only condition is that the interaction Lagrangian is a Lorentz-invariant polynomial in the fields and their derivatives. In particular, operators of any dimension in the UV theory and EFT can be included.

We first overview here the methods used internally by \texttt{MatchingTools}. Given a UV theory defined by an action $S_{\text{UV}}[\phi, \Phi]$, with light fields $\phi$ and heavy fields $\Phi$, \texttt{MatchingTools} integrates out $\Phi$ at tree level by solving its equation of motion and replacing it in $S_{\text{UV}}$. That is, the effective action is given by
\begin{equation}
  S_{\text{EFT}}[\phi] = S_{\text{UV}}[\phi, \Phi_c(\phi)],
  \qquad \qquad
  \text{where }
  \left.\frac{\delta S}{\delta \Phi}\right|_{\Phi = \Phi_c(\phi)} = 0.
\end{equation}
\texttt{MatchingTools} computes the solution $\Phi_c(\phi)$ as a perturbative expansion in inverse powers of the mass $M$ of $\Phi$. It does so by means of an iterative procedure that generates a sequence of solutions $\Phi_n(\phi)$, starting with $\Phi_0(\phi) \equiv 0$ and given by
\begin{equation}
  \Phi_n(\phi) \equiv \left. P \frac{\delta S_{\text{int}}}{\delta \Phi} \right|_{\Phi = \Phi_{n - 1}(\phi)},
\end{equation}
where $P$ is the propagator for $\Phi$, expanded in powers of $1/M$, and $S_{\text{int}}[\phi, \Phi]$ is the interaction part of the UV action.
Each $\Phi_n(\phi)$ is a solution of the equations of motion only to a finite order in $1/M$, but this order increases with $n$. \texttt{MatchingTools} can thus iterate this procedure to compute the solution to any order in $1/M$, which in turn gives the effective Lagrangian to any desired order.

The effective Lagrangian obtained from this method will contain in general a set of operators that are not independent. In order to re-write it in terms of a set of independent operators, a basis, three different types of operations can be applied to it: algebraic/group theory identities, field redefinitions (or, equivalently at leading order, using equations of motion), and integration by parts. \texttt{MatchingTools} unifies all of them under a general system for finding and replacing patterns in the effective Lagrangian. The patterns that can be replaced are products of fields and constant tensors, with arbitrary index contractions.

We now consider a simple example to illustrate the usage and features of \texttt{MatchingTools}. The UV theory has a $SU(2) \times U(1)$ symmetry, and it contains two scalar multiplets: $\phi$, a light doublet with hypercharge 1/2; and $\Phi$, a heavy triplet with vanishing hypercharge. Their interactions are given by
\begin{equation}
  \mathcal{L}_{\text{int}} \supset
  - \kappa \, \Xi^a (\phi^\dagger \sigma^a \phi)
  - \lambda \, (\Xi^a \Xi^a) (\phi^\dagger \phi).
\end{equation}
This theory can be defined in \texttt{MatchingTools} using the following code:
\begin{lstlisting}[language=iPython]
import matchingtools as mt

sigma = mt.TensorBuilder("sigma")
kappa = mt.TensorBuilder("kappa")
lamb = mt.TensorBuilder("lamb")

phi = mt.FieldBuilder("phi", 1, mt.boson)
phic = mt.FieldBuilder("phic", 1, mt.boson)
Xi = mt.FieldBuilder("Xi", 1, mt.boson)

L_int = -mt.OpSum(
  mt.Op(kappa(), Xi(0), phic(1), sigma(0, 1, 2), phi(2)),
  mt.Op(lamb(), Xi(0), Xi(0), phic(1), phi(1)),
)
\end{lstlisting}
First, the different symbols that appear in the Lagrangian are defined: the Pauli matrices $\sigma$, the coupling constants $\kappa$ and $\lambda$, and the fields $\phi$ and $\Xi$. All these objects are viewed by \texttt{MatchingTools} as tensors, possibly with zero indices, as in the case of the coupling constants in this example. For the fields, their canonical dimension and commutation properties have to be specified. Finally, the interaction Lagrangian is constructed as a sum (\texttt{OpSum}) of operators (\texttt{Op}). Each operator is given by the list of its factors, which can be both fields and constant tensors. The index structure of the operator is expressed by placing a non-negative integer in each position corresponding to an index, with repeated integers denoting contraction.

The program is now ready to integrate out $\Xi$. To do so, one can write:
\begin{lstlisting}[language=iPython,firstnumber=15]
heavy_Xi = mt.RealScalar("Xi", 1, has_flavor=False)
L_eff = mt.integrate(
  heavy_fields=[heavy_Xi], interaction_lagrangian=L_int, max_dim=6
)
print(mt.Writer(L_eff, []))
\end{lstlisting}
This produces a list of all the terms in the resulting effective Lagrangian. To re-write it in terms of a basis of operators, one can make use of the find-and-replace system provided by \texttt{MatchingTools}. As an example, we will use the $SU(2)$ Fierz identity:
\begin{equation}
  \sigma^a_{ij} \sigma^a_{kl} = 2 \delta_{il} \delta_{jk} - \delta_{ij} \delta_{kl},
\end{equation}
to simplify the Lagrangian, by replacing every occurrence of the left-hand side of this equation by its right-hand side. To do this, we define the corresponding rule, which is a tuple whose first element is the pattern and whose second element is the replacement. We then apply this rule to the effective Lagrangian:
\begin{lstlisting}[language=iPython,firstnumber=20]
fierz_rule = (
  mt.Op(sigma(0, -1, -2), sigma(0, -3, -4)),
  mt.OpSum(
    number_op(2) * mt.Op(mt.kdelta(-1, -4), mt.kdelta(-3, -2)),
    -mt.Op(mt.kdelta(-1, -2), mt.kdelta(-3, -4))
  )
)
L_eff = mt.apply_rules(L_eff, [fierz_rule], max_iterations=1)
print(mt.Writer(mt.simplify(L_eff), []))
\end{lstlisting}
This code outputs the list of terms of the transformed Lagrangian. After all redundant terms have been removed and the Lagrangian is written as a linear combination of operators in the desired basis, the final step is usually to identify the coefficients of the operators in the basis. To do this, one can define rules to replace the explicit expression of each operator by a single symbol and then instruct \texttt{MatchingTools} to extract the coefficients of these symbols. For the purpose of our example, we do so for an overcomplete set operators, since we have not reduced all redundancies yet. The overcomplete basis is:
\begin{align}
  \mathcal{O}_{\phi 6} &= (\phi^\dagger\phi)^3, &
    \mathcal{O}_{\phi 4} &= (\phi^\dagger\phi)^2, \\
    \mathcal{O}^{(1)}_{\phi} &= \phi^\dagger\phi
    (D_\mu \phi)^\dagger D^\mu \phi, &
    \mathcal{O}^{(3)}_{\phi}&= (\phi^\dagger D_\mu \phi)
    (D^\mu \phi)^\dagger \phi, \\
    \mathcal{O}_{D \phi} &= \phi^\dagger(D_\mu \phi)
    \phi^\dagger D^\mu\phi, &
    \mathcal{O}^*_{D \phi} &= (D_\mu\phi)^\dagger\phi
    (D^\mu\phi)^\dagger\phi,
\end{align}
and the code to obtain the corresponding coefficients:
\begin{lstlisting}[language=iPython,firstnumber=29]
Ophi6 = mt.tensor_op("Ophi6")
Ophi4 = mt.tensor_op("Ophi4")
...

definition_rules = [
  (mt.Op(phic(0), phi(0), phic(1), phi(1), phic(2), phi(2)), mt.OpSum(Ophi6)),
  (mt.Op(phic(0), phi(0), phic(1), phi(1)), mt.OpSum(Ophi4)),
  (mt.Op(mt.D(2, phic(0)), mt.D(2, phi(0)), phic(1), phi(1)), mt.OpSum(O1phi)),
  (mt.Op(phic(0), mt.D(2, phi(0)), mt.D(2, phic(1)), phi(1)), mt.OpSum(O3phi)),
  (mt.Op(phic(0), mt.D(2, phi(0)), phic(1), mt.D(2, phi(1))), mt.OpSum(ODphi)),
  (mt.Op(mt.D(2, phic(0)), phi(0), mt.D(2, phic(1)), phi(1)), mt.OpSum(ODphic))
]

L_eff = mt.apply_rules(L_eff, definition_rules, 1)
final_coef_names = ["Ophi6", "Ophi4", "O1phi", "O3phi", "ODphi", "ODphic"]
print(mt.Writer(L_eff, final_coef_names))
\end{lstlisting}
With output:
\begin{verbatim}
  O1phi:
    2 (MXi^(-4)) kappa kappa
  O3phi:
    -1 (MXi^(-4)) kappa kappa
  ...
\end{verbatim}
Indicating that the coefficient of the operator $\mathcal{O}^{(1)}_{\phi}$ is $2 \kappa^2 / M_{\Xi}^4$, the coefficient of $\mathcal{O}^{(3)}_{\phi}$ is $-\kappa^2 / M_{\Xi}^4$, \ldots This can be converted to LaTeX code using the method \texttt{write\_latex} from the \texttt{Writer} class.

\texttt{MatchingTools} also provides an \texttt{extras} subpackage containing modules for SMEFT-related applications, including the definition of tensors and rules relevant for $SU(2)$, $SU(3)$ and Lorentz group theory, the definitions of the SM fields, the rules for applying the SM equations of motion, and the definitions of the Warsaw basis operators. More information on these modules and other \texttt{MatchingTools} features can be found at \href{https://matchingtools.readthedocs.io}{matchingtools.readthedocs.io}.

\subsection{{\tt STrEAM}}

\STrEAM\ (\textbf{S}uper\textbf{Tr}ace \textbf{E}valuation \textbf{A}utomated for \textbf{M}atching) is a Mathematica\ package that automates the evaluation of functional supertraces that could arise when one matches a generic UV theory onto a relativistic EFT. \STrEAM\ implements the covariant derivative expansion method and could provide the result to arbitrary order in the heavy mass expansion.

According to the streamlined functional matching prescription presented in Ref.~\cite{Cohen:2020fcu}, the matching result at the one-loop level can be computed by evaluating the functional supertraces
\begin{equation}
\int \text{d}^4 x \,\mathcal{L}_\text{EFT}^\text{(1-loop)}[\phi] = \frac{i}{2}\,\STr\log \bm{K} \Bigr|_\text{hard}
- \frac{i}{2} \sum_{n=1}^\infty \frac{1}{n}\, \STr\Big[ \big( \bm{K}^{-1} \bm{X} \big)^n \Big] \Bigr|_\text{hard} \,,
\label{eqn:SuperTraces}
\end{equation}
where the inverse (covariant) propagator matrix $\bm{K}$ is diagonal
\begin{equation}
K_i =
\begin{cases}
P^2 - m_i^2 & \big(\text{spin-}0 \big) \\[3pt]
\Psl - m_i  & \big(\text{spin-}\frac{1}{2}\big) \\[3pt]
-\eta^{\mu\nu} (P^2 - m_i^2) & \big(\text{spin-}1\big)
\end{cases}\quad,
\label{eqn:Ki}
\end{equation}
and the interaction matrix $\bm{X}$ can be organized into a derivative expansion:
\begin{equation}
\bm{X}(\phi, P_\mu) = \bm{U}[\phi] + \bigl(P_\mu \bm{Z}^\mu[\phi] +\bm{\bar{Z}}^\mu[\phi] P_\mu\bigr) + \cdots \,,
\label{eqn:bmX}
\end{equation}
with $P_\mu \equiv iD_\mu$ the ``open'' covariant derivative. Therefore, a general power-type supertrace in Eq.~\eqref{eqn:SuperTraces}
\begin{equation}
-i\, \STr \bigg[ \frac{1}{K_{i_1}} X_{i_1 i_2}\, \frac{1}{K_{i_2}} X_{i_2 i_3}\, \cdots\, \frac{1}{K_{i_n}} X_{i_n i_1} \bigg] \,,
\label{eqn:KXstring}
\end{equation}
consists of a product sequence of segments of the form
\begin{equation}
\frac{1}{K_i}\, \big( P_{\mu_1} \cdots P_{\mu_n} \big)\, U_k\, \big( P_{\nu_1} \cdots P_{\nu_m} \big) \,.
\label{eqn:segments}
\end{equation}

Using $\Delta_i$ and $\Lambda_i$ to denote the bosonic and fermionic versions of $K_i^{-1}$, respectively
\begin{equation}
\Delta_i \equiv \frac{1}{P^2-m_i^2} \,,\qquad
\Lambda_i \equiv \frac{1}{\Psl-m_i} \,,
\label{eqn:DLdef}
\end{equation}
the concrete scope of \STrEAM\ can be summarized as following:
\vspace{1mm}
\begin{tcolorbox}[colback=light-gray]
\begin{center}
\begin{minipage}{5.5in}
\STrEAM\ automates the evaluation of functional supertraces of the form
\begin{equation}
-i\,\STr \Big[\, f \big( P_\mu \,,\, \left\{U_k\right\} \big) \Big] \Bigr|_\text{hard} \,,
\label{eqn:STrfh}
\end{equation}
where $f$ is a product sequence of $P_\mu$, $U_k$, $\Delta_i$ and $\Lambda_i$, consisting of an arbitrary number of ``propagator blocks'':
\begin{equation}
f \,=\, \Big[ \,\,\cdots\,\, \big( P_{\mu_1} \dots P_{\mu_n} \big)\, \big(\, \Delta_i \,\text{ or }\, \Lambda_i\, \big)\, \big( P_{\nu_1} \dots P_{\nu_m} \big)\, U_k \,\, \cdots\,\,\Big] \,.
\label{eqn:fSTrEAM}
\end{equation}
\end{minipage}
\end{center}
\end{tcolorbox}
\vspace{2mm}\noindent
The last block in $f$ is allowed to have a trivial $U$ factor, \ie\ $U=1$, such that the log-type supertraces in Eq.~\eqref{eqn:SuperTraces} can also be covered upon taking a mass derivative
\begin{subequations}
\begin{align}
\frac{\partial}{\partial m_\Phi^2} \Big[ i\,\STr\log\left(P^2 - m_\Phi^2\right) \Big] &= -i\,\STr \bigg[ \frac{1}{P^2 - m_\Phi^2} \bigg] = -i\,\STr \big[ \Delta_\Phi \big] \bigr|_\text{hard} \,, \\[5pt]
\frac{\partial}{\partial m_\Phi} \Big[ i\,\STr\log\left(\Psl - m_\Phi \right) \Big] &= -i\,\STr \bigg[ \frac{1}{\Psl - m_\Phi} \bigg] = -i\,\STr \big[ \Lambda_\Phi \big] \bigr|_\text{hard} \,.
\end{align}
\end{subequations}

The \STrEAM\ package can be downloaded from GitHub at
\begin{center}
\url{https://www.github.com/EFTMatching/STrEAM}
\end{center}
After placing the file ``STrEAM.m'' at the user's own choice of directory ``\texttt{/path/to/package/}'', one can load it with the usual Mathematica\ command:
\begin{mmaCell}[index=1]{Input}
<<"/path/to/package/STrEAM.m";
\end{mmaCell}
\STrEAM\ is a compact package with a single main function \mmaInlineCell{Code}{SuperTrace}. It has a simple syntax:
\begin{mmaCell}{Input}
\mmaDef{SuperTrace}[dim, flist]
\end{mmaCell}
with two mandatory arguments: \mmaInlineCell{Code}{dim} is an \mmaInlineCell{Code}{Integer} that specifies the desired operator dimension in the evaluation result; \mmaInlineCell{Code}{flist} is a \mmaInlineCell{Code}{List} that specifies the functional operator $f\big( P_\mu, \left\{U_k\right\} \big)$ to be traced over; it consists of \mmaInlineCell{Input}{\mmaSub{P}{\(\mmaUnd{\mu}\)}}, \mmaInlineCell{Input}{\mmaSub{U}{k}}, \mmaInlineCell{Input}{\mmaSub{\(\mmaUnd{\Delta}\)}{i}}, and \mmaInlineCell{Input}{\mmaSub{\(\mmaUnd{\Lambda}\)}{i}}, organized in the form of Eq.~\eqref{eqn:fSTrEAM}. The main function \mmaInlineCell{Code}{SuperTrace} also has a few options; see Sec. 4 in Ref.~\cite{Cohen:2020qvb} for a list of them.

As a simple demonstration example, the result
\begin{align}
-i\,\STr \bigg[ \frac{1}{P^2-m_1^2} U_1^{[2]} \bigg] \biggr|_\text{hard} = \int\text{d}^4x \,\tfrac{1}{16\pi^2}\, \text{tr} \bigg[ m_1^2 \Big( 1 - \log\tfrac{m_1^2}{\mu^2} \Big)\, U_1 + \tfrac{1}{12 m_1^2}\, F_{\mu\nu} F^{\mu\nu} U_1 \bigg] \,,
\end{align}
can be obtained by calling \mmaInlineCell{Code}{SuperTrace} as
\begin{mmaCell}[index=3]{Input}
\mmaDef{SuperTrace}[6, \{\mmaSub{\(\Delta\)}{1}, \mmaSub{U}{1}\}, Udimlist->\{2\}, display->True];
\end{mmaCell}
which will print
\begin{mmaCell}{Print}
-iSTr[\mmaFrac{1}{\mmaSup{P}{2}-\mmaSubSup{m}{1}{2}}\mmaSub{U}{1}]\mmaSub{|}{hard} = \(\int\)\mmaSup{d}{4}x \mmaFrac{1}{16\mmaSup{\(\pi\)}{2}} tr\{

\(\qquad\) \mmaSubSup{m}{1}{2}\bigg(1-Log\Big[\mmaFrac{\mmaSubSup{m}{1}{2}}{\mmaSup{\(\mu\)}{2}}\Big]\bigg) \(\qquad\) (\mmaSub{U}{1}) \(\qquad\qquad\qquad\qquad\,\) (dim-2)

\(\qquad\) \mmaFrac{1}{12\mmaSubSup{m}{1}{2}} \(\qquad\qquad\qquad\;\;\;\) (\mmaSub{F}{\mmaSub{\(\mu\)}{1}\mmaSub{\(\mu\)}{2}})(\mmaSub{F}{\mmaSub{\(\mu\)}{1}\mmaSub{\(\mu\)}{2}})(\mmaSub{U}{1}) \(\qquad\) (dim-6)

\}
\end{mmaCell}
With the option \mmaInlineCell{Input}{display->True}, \mmaInlineCell{Code}{SuperTrace} will print the evaluation result in \mmaInlineCell{Code}{TableForm}, together with the input supertrace, as shown above. More demonstration examples, as well as a more detailed manual of \STrEAM\ can be found in Sec. 4 of Ref.~\cite{Cohen:2020qvb}.

\subsection{General procedure for code comparison}

A key step in the development of automated matching tools is cross-validation and comparison between different theoretical approaches and code implementations. Indeed, a variety of cross-checks have already been implemented for the different matching codes. For instance, the CDE of multiple supertraces has been validated by direct comparison of {\tt Supertracer} and {\tt STrEAM} outputs. Moreover, both {\tt MatchmakerEFT} and {\tt Supertracer} have already been used and partially cross-checked in the context of specific UV models~\cite{Gherardi:2020det,Chala:2020wvs,Chala:2021pll,Chala:2021wpj,Chala:2021cgt,Zhang:2021jdf,Dedes:2021abc,Crivellin:2022fdf,BakshiDas:2022xhd,Guedes:2022cfy,Du:2022vso,Li:2022ipc}. CoDEx generated matching results for sixteen SM extensions with a single heavy scalar are available here \href{https://github.com/effExTeam/Precision-Observables-and-Higgs-Signals-Effective-passageto-select-BSM}{\faGithub}. SILH basis matching results are cross-checked with the models given in Ref.~\cite{Henning:2014wua} and they agree. Warsaw basis matching result for singlet real scalar extensions of the SM is cross-checked with Refs.~\cite{Jiang:2018pbd,Haisch:2020ahr} and it agrees. However, as matching codes become more mature, it becomes highly desirable to have more systematic and comprehensive cross-checks. 

In an effort to establish a well-defined standard for cross-validation, we have created the GitLab repository \url{https://gitlab.com/modelmatch/ModelMatch}. This repository will be used as an archive for BSM to SMEFT (and possibly other EFTs down the line) matching calculations, at the time that it will provide a transparent and open-access framework for comparison among the different implementations. To this end, any matching calculation presented in this repository will contain the following three files:\footnote{More details will be provided in the GitLab repository with some basic information also publically available at \url{https://twiki.cern.ch/twiki/bin/view/LHCPhysics/EFTAC5}.}
\begin{enumerate}
    \item {\bf Matching results:} in any format that the authors deem appropriate.
    \item {\bf Validation:} consisting of a WCxf file with numerical matching coefficients for a given set of benchmark parameters for comparison with other implementations.
    \item {\bf Additional information:} provided in the form of a document clearly stating (at least) the following information:
    \begin{itemize}
        \item Corresponding author(s).
    
        \item All theory assumptions entering into the one-loop matching computation, including renormalization scheme, $\gamma_5$ prescription, gauge-fixing procedure, metric signature, and Levi-Civita convention. 
    
        \item The complete UV Lagrangian. In case of heavy vectors, an additional Lagrangian in the broken phase is highly encouraged. 
    
        \item The set of benchmark parameter values used in the validation file. To avoid possible numerical issues, factorizing the loop factors and using rational values for the model parameters would be preferred. 
    \end{itemize}
\end{enumerate}

We propose the following representative examples of BSM models including, respectively, heavy scalars, fermions, and vectors: $S_1+ S_3$ scalar leptoquark extension, a heavy vector-like lepton transforming under the SM gauge group as $E\sim(\rep{1},\rep{1},-1)$, and a heavy vector triplet from the symmetry breaking $ \SU(2)'_\LL \times \SU(2)_X \to \SU(2)_\LL$. As an ultimate test for the long-term future, we will also consider the matching of the SMEFT to the Low-Energy Effective Field Theory (LEFT), where the Higgs, top, $W$, and $Z$ are integrated out. This latter example is particularly comprehensive as it involves simultaneously heavy scalars, fermions, and vectors.

\subsection{Outlook}

The dream is that one could take a Lagrangian (e.g. implemented in FeynRules) and then pass it through a code that spits out the 1-loop Wilson coefficients in the Warsaw basis automatically.  Putting all of this together is an area of active interest. The field will surely move forwards by leaps and bounds, once the next generation of automated matching tools becomes operational at which point cross-validation of the codes will become very important.

\section{Supplementary numerical Codes}
\label{sec:NumCodes}
In this section, we discuss additional codes that are used in phenomenological analyses for SMEFT and LEFT running or to compare different matching results.

\subsection{\texttt{DsixTools}}

\texttt{DsixTools}~\cite{Celis:2017hod,Fuentes-Martin:2020zaz} is an open-source Mathematica package that automates one-loop RGE in the SMEFT~\cite{Jenkins:2013zja,Jenkins:2013wua,Alonso:2013hga,Alonso:2014zka} and in the LEFT~\cite{Jenkins:2017dyc}, as well as one-loop SMEFT-to-LEFT matching~\cite{Jenkins:2017jig,Dekens:2019ept,Aebischer:2015fzz}. One of the main features of \texttt{DsixTools} is that it contains not only numerical but also analytical routines, allowing for simple manipulation of beta functions and matching expressions. 

\texttt{DsixTools 2.1} aims for a more visual and user-friendly experience. Together with the usual matching and running routines, it also provides an interface with useful information for all operators and parameters of the SMEFT and the LEFT, such as their flavor symmetries and the number of degrees of freedom. Routines for the implementation of this information on global expressions (like decay amplitudes and cross-sections) are also provided. Furthermore, \texttt{DsixTools} includes a user-friendly input that performs automatic consistency checks, simplifying the user’s task.

The package can be simply installed by running the following command in a Mathematica notebook (it is advised to use a fresh kernel for the installation):
\mmaSet{index=1}
\begin{mmaCell}{Input}
Import["https://raw.githubusercontent.com/DsixTools/DsixTools/master/install.m"]
\end{mmaCell}
This will download and install \texttt{DsixTools} in the Applications folder of the Mathematica base directory. It will also create Mathematica documentation for all the package routines.

The following lines provide a basic (yet complete) usage example:
\begin{mmaCell}{Input}
Needs["DsixTools\`{}"]
\mmaDef{NewScale}[\{\mmaDef{HIGHSCALE} -> 10000\}]; 
\mmaDef{NewInput}[\{\mmaDef{Clq1}[1 ,1 ,1 ,2] -> 1/\mmaDef{HIGHSCALE}^2, \mmaDef{Clq1}[1 ,1 ,2 ,1] -> 1/\mmaDef{HIGHSCALE}^2, \mmaDef{CH} -> -0.5/\mmaDef{HIGHSCALE}^2\}]; 
\mmaDef{RunDsixTools}; 
\mmaDef{D6run}[\mmaDef{LeuVLL}[2 ,2 ,1 ,1]] /. \textbackslash[\mmaDef{Mu}] -> \mmaDef{LOWSCALE}
\end{mmaCell}
which illustrates how to load the package, provide the input for a given SMEFT Lagrangian at the UV scale $\Lambda_{\rm UV} =$ \texttt{HIGHSCALE} (in units of GeV), and calculate the LEFT WCs at the IR scale $\Lambda_{\rm IR} =$ \texttt{LOWSCALE} (by default, \texttt{LOWSCALE} = 5~GeV). 

\texttt{DsixTools} also offers multiple options to interface with other EFT tools. In particular, it can import and export JSON and YAML files in the \texttt{WCxf} exchange format~\cite{Aebischer:2017ugx} (see Sec.~\ref{sec:WCxf} for further details). It also admits as an input the output generated by \texttt{Matchmakereft} (see Sec.~\ref{sec:MMEFT}). For instance, a \texttt{Matchmakereft} output file (MMEfile.dat) is loaded into \texttt{DsixTools} using the command line:
\begin{mmaCell}{Input}
\mmaDef{NewInput}[\{\mmaDef{MMEfile} -> "MMEfile.dat " , MS -> 1000 , lam2 -> 0.1 , lam3 -> 0.2\}, \mmaDef{HIGHSCALE} -> 1000]
\end{mmaCell}
where all the  UV-model parameters (in this example MS, lam2 and lam3) and the \texttt{HIGHSCALE} must be assigned numerical values.

Additional information and an up-to-date version of the user's manual can be found on the package webpage: \url{https://dsixtools.github.io/}, or in the project's \href{https://github.com/DsixTools/DsixTools}{GitHub repository}.

\subsection{\texttt{RGESolver}}

\texttt{RGESolver}~\cite{DiNoi:2022ejg} is an open-source \texttt{C++} library
that performs the renormalization group evolution of the SMEFT Wilson coefficients in a fast and easy-to-use manner. The library deals with the most generic flavor scenario, assuming only lepton and baryon number conservation. 

\texttt{RGESolver} has been developed with a specific focus on extensive phenomenological analyses. Two methods are available in order to solve the differential equations: a numerical solution or an approximate one (first leading log). Furthermore, after the RGE \texttt{RGESolver} can perform the so-called flavor back-rotation \cite{Aebischer:2020lsx}. It also provides a routine to perform the evolution and the back-rotation with a simple command.
\texttt{RGESolver} has been tested against \texttt{DSixTools 2.1} \cite{Celis:2017hod,Fuentes-Martin:2020zaz} for both the numerical and the approximate method, obtaining differences between the two codes $\lesssim 10^{-5}\, 1/\Lambda^2$ for initial conditions $\mathcal{O}(1/\Lambda^2)$.

\texttt{RGESolver} can generate initial conditions for the Standard Model parameters (gauge couplings, Higgs sector parameters, and Yukawa matrices) at any given scale solving pure Standard Model renormalization group equations. 

We give a simple example: these few lines generate the initial conditions (in the basis where the down Yukawa matrix is diagonal) for the Standard Model parameters at the scale $\mu = \Lambda=10$ TeV, set $\mathcal{C}^{dH}_{1,2}(\Lambda)=1/\Lambda^2$, solve numerically the renormalization group equations down to $\mu = 250$ GeV and perform the back-rotation to get back into the original basis. Finally, the real part of the evolved coefficient $\mathcal{C}^{dH}_{1,2}(250\,\textrm{GeV})$ can be accessed via the dedicated getter method.
\begin{verbatim}
double Lambda = 10000.;
S.GenerateSMInitialConditions(Lambda, "DOWN", "Numeric");
S.SetCoefficient("CHdR", 1. / (Lambda * Lambda), 1, 2);
S.EvolveToBasis("Numeric", Lambda, 250., "DOWN");
double EvolvedCHdR_12 = S.GetCoefficient("CHdR", 1, 2);
\end{verbatim}
All the details about the installation, the extended documentation, and some examples can be found in the \href{https://github.com/silvest/RGESolver}{\texttt{GitHub} dedicated page}. 

\subsection{\texttt{WCxf}}\label{sec:WCxf}

The Wilson coefficient exchange format (\texttt{WCxf}) defines a standard for Wilson coefficients used in computer codes. Since many different conventions are being used in the literature, it is important to have a collection of unique definitions for the different bases and the corresponding Wilson coefficients, especially when comparisons between different codes are performed. A list of all the computer tools that already support \texttt{WCxf} as well as their corresponding bases can be found on the \texttt{WCxf} GitHub website \url{https://wcxf.github.io/}.
The format is extensible in the sense that any new basis can be added to the predefined bases, by providing simply a yaml file in which the convention for the Wilson coefficients together with the underlying EFT are specified. Further details on how to extend \texttt{WCxf} can be found in~\cite{Aebischer:2017ugx} and on the GitHub webpage.

Furthermore, there is the Python module \texttt{WCxf}, which allows changing numerical values of Wilson coefficients between different operator bases. Especially when comparing results obtained with different computer codes this module comes in handy, as it allows translation of the Wilson coefficients from one code in an automated way into the basis of the other, which facilitates cross-checks and comparisons. In the following we will show how to translate Wilson coefficients from one basis into another using \texttt{WCxf}:

The package can be easily installed, using

\begin{lstlisting}[language=iPython]
python3 -m pip install wcxf --user
\end{lstlisting}

Importing a given \texttt{WCxf} yaml file that specifies the Wilson coefficient values is done by:

\begin{lstlisting}[language=iPython]
import wcxf

with open('my_wcxf_input_file.yml', 'r') as f:
  wc = wcxf.WC.load(f)
\end{lstlisting}

Translating the Wilson coefficients of the imported \texttt{WCxf} file into another basis is achieved by

\begin{lstlisting}[language=iPython]
wc_new = wc.translate('My target basis')
\end{lstlisting}

Further details on \texttt{WCxf} as well as further commands can be found in \cite{Aebischer:2017ugx}.

\subsection{\texttt{wilson}}
The Python package \texttt{wilson}~\cite{Aebischer:2018bkb} is a matchrunning tool, which is mainly used in phenomenological codes such as {\tt flavio}\cite{Straub:2018kue} and {\tt smelli}\cite{Aebischer:2018iyb}, but can also be used independently. It allows to numerically run all Wilson coefficients in the SMEFT to arbitrary scales, as well as matching them onto the LEFT. Furthermore, the full LEFT running below the electroweak scale is implemented in \texttt{wilson}.
The implementation entails
\begin{itemize}
\item The complete one-loop SMEFT beta functions \cite{Alonso:2013hga, Jenkins:2013zja, Jenkins:2013wua}.
\item The complete tree-level \cite{Aebischer:2015fzz, Jenkins:2017jig} and one-loop \cite{Dekens:2019ept} matching from SMEFT onto LEFT.
\item The complete one-loop LEFT running \cite{Aebischer:2017gaw, Jenkins:2017dyc}.
\end{itemize}

\texttt{wilson} also supports the \texttt{WCxf} format \cite{Aebischer:2017ugx} and takes back-rotation \cite{Aebischer:2020lsx} effects into account, which result when mass matrices have to be rediagonalized after RGE running.

The package can be installed using

\begin{lstlisting}[language=iPython]
python3 -m pip install wilson --user
\end{lstlisting}

To set a Wilson coefficient to a certain value at a particular scale in a given EFT the Wilson class is used. For instance, setting the Wilson coefficient of the SMEFT operator $\mathcal{O}^{23}_{dG}=\left( \bar q_2 \sigma^{\mu \nu} T^A d_3 \right) \varphi \,G_{\mu \nu}^A$ to one at the scale $\Lambda=1\,\text{TeV}$ in the Warsaw basis one writes:

\begin{lstlisting}[language=iPython]
from wilson import Wilson
mywilson = Wilson({'dG_23': 1e-6}, scale=1e3, eft='SMEFT', basis='Warsaw')
\end{lstlisting}

This Wilson coefficient can then be run down to the EW scale, matched onto the Weak Effective Theory (WET), which is equivalent to the LEFT, and further run down to a lower scale. For example, matching onto the JMS basis (introduced in \cite{Jenkins:2017jig}) and running to $100\,\text{GeV}$ is achieved by

\begin{lstlisting}[language=iPython]
wc_JMS = mywilson.match_run(scale=100, eft='WET', basis='JMS')
\end{lstlisting}

Further information and updates can be found on the project website \url{https://wilson-eft.github.io/} and in \cite{Aebischer:2018bkb}.

\section{Conclusions and Outlook}
\label{sec:conclusions}

Several codes aiming towards a full automatization of matching the short-distance BSM models to the SMEFT have been presented alongside some numerical codes for the basis conversion and the low-energy matching and running. 
A procedure to cross-validate the different matching approaches has been suggested for future work, identifying several interesting models.
\vspace{5pt}

\!\!\!\!\!\!\!\!\!\!\!\!\!\!{\bf Acknowledgments}: This work was done on behalf of the LHC EFT WG and we would like to thank members of the LHC EFT WG for stimulating discussions which led to this document.

\bibliographystyle{JHEP}
\bibliography{bib}

\end{document}